# Security and Privacy of performing Data Analytics in the cloud – A three-way handshake of Technology, Policy and Management

[1]Nidhi Rastogi, [2]Marie Joan Kristine Gloria and [3]Dr. James Hendler

[1]raston@rpi.edu, [2]glorim@rpi.edu, [3]hendler@cs.rpi.edu

**ABSTRACT**

Cloud platform came into existence primarily to accelerate IT delivery and to promote innovation. To this point, it has performed largely well to the expectations of technologists, businesses and customers. The service aspect of this technology has paved the road for a faster set up of infrastructure and related goals for both startups and established organizations. This has further led to quicker delivery of many user-friendly applications to the market while proving to be a commercially viable option to companies with limited resources. On the technology front, the creation and adoption of this ecosystem has allowed easy collection of massive data from various sources at one place, where the "place" is sometimes referred to as just the cloud. Efficient data mining can be performed on raw data to extract potentially useful information, which was not possible at this scale before. Targeted advertising is a common example that can help businesses. Despite these promising offerings, concerns around security and privacy of user information suppressed wider acceptance and an all-encompassing deployment of the cloud platform.

In this paper, we discuss security and privacy concerns that occur due to data exchanging hands between a cloud servicer provider (CSP) and the primary cloud user – the data collector, from the content generator, see Figure 1. We offer solutions that encompass technology, policy and sound management of the cloud service asserting that this approach has the potential to provide a holistic solution.

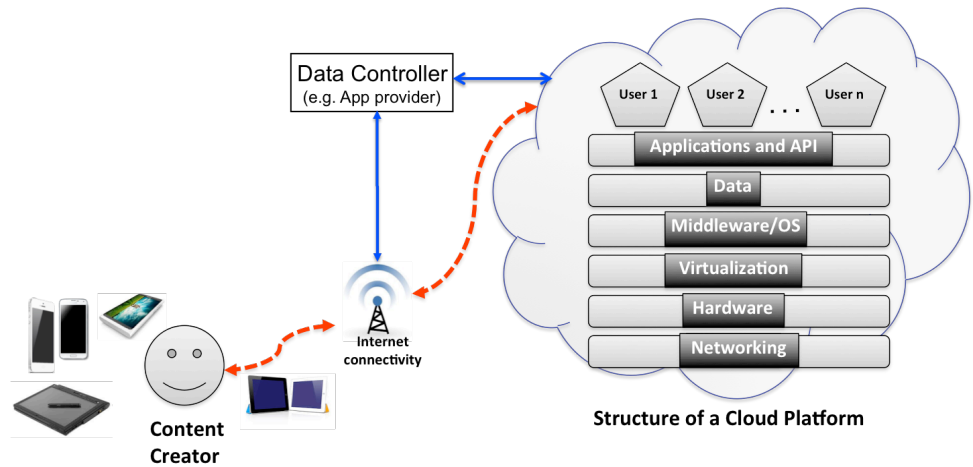

Figure 1- The cloud ecosystem

## INTRODUCTION

Cloud-computing platform offers opportunities for developers to deploy mobile applications dynamically on a scalable on-demand hardware and software platform. It includes some unique features such as a complete end-to-end infrastructural solution with enough computation and storage resources as well as no maintenance responsibilities. All these features come considering the need for economies of scale by parties who wouldn't be able to afford them otherwise. Companies like Salesforce, Oracle, Amazon, Google, and IBM have found this model lucrative and have created a cloud division within their respective organizations. Machine virtualization techniques have been deployed to provide flexible and cost-effective resource sharing for users both internal and external to the organization. This has encouraged individual developers and small size companies, like Dropbox, to create cloud platform orientated services and products that are interesting to the end user.

However, data privacy concerns have thwarted the pace of its deployment. The user, who entrusts the CSP with personal data, is also expected extend this trust to third parties on matters related to its access. The platform thus acts like a "black-box" where the cloud service provider (CSP) is largely in control of gigabytes of user information. This information can range from highly sensitive to publicly available. Concerns are raised when parties that are interested in user-data analytics deploy artificial intelligence (AI) techniques, including machine-learning algorithms, to identify targeted audience for various purposes; advertisements being one of them. This has negatively influenced the mindset of data owners who are provided with no guarantees by the CSPs that completely prohibit further usage of their data by anyone. Hence, it is difficult for the consumer to believe that the service provider will not share data covertly to a party outside of the original usage agreement.

A strong enabler for preventing unauthorized access of information is encryption. It encodes data of all types into a format that is readable only to authorized parties. A suitable solution for this environment among all encryptions that serves the purpose of maintaining data privacy is homomorphic encryption[1]. Another similar mechanism called parallel homomorphic encryption (PHE)[2] supports intensive computations via evaluation algorithms that can be efficiently executed in parallel. Encryption allows computation on encrypted data within the cloud without having the need to decrypt it, thus preventing exposure to those who have no legitimate need for data access. However, like many other strong encryption schemes, these protocols come with additional computational overhead of working on encrypted information. Although PHE is an improvement over homomorphic encryption in terms of faster computation, just like the homomorphic encryption, a lot of work is required to make it viable on a commercial scale.

While a lot of technical research worth mentioning has been going on towards realizing a feasible full-proof solution for preventing unauthorized data access, we redirect the reader's attention towards alternate ways of dealing with the problem. In this Article, we suggest a three-pronged strategy to get a grip on this situation driven by– a) Technology, b) Internal Policy and Management, and c) State and Federal Policies. Our recommendation considers that while technology is a powerful agent in preserving data confidentiality in a cloud setup, it is insufficient in providing a complete solution unless backed by appropriate practices. A sound privacy assessment of the cloud also requires transparent pro-user management practices and internal policies such as: a) softwares that manage low-risk data cohabitate with those that have similar security needs; b) a blueprint of threat modeling of the cloud service - including software, hardware, and data; and c) a mechanism that addresses accountability concerns for protecting all data and control information that is used to grant access to the various parties. Lastly, we call for further exploration of external policies on both the state and federal level that offer limits and safeguards for the entire ecosystem. We submit these ideas in hopes to generate interest among policy makers, technologists, researchers and industry to consider as potential practical steps towards better data management.

---

[1] Gentry, Craig. "Fully homomorphic encryption using ideal lattices." In STOC, vol. 9, pp. 169-178. 2009.
[2] Kamara, Seny, and Mariana Raykova. "Parallel Homomorphic Encryption." In Financial Cryptography and Data Security, pp. 213-225. Springer Berlin Heidelberg, 2013.

**LITERATURE REVIEW**

A brief literature overview of the existing privacy and security concerns related to cloud platform is covered below. For this Article, we define the various players in the ecosystem as follows: A cloud service is rendered over a network and can be accessed remotely through the Internet. A cloud service provider (CSP) is the entity that provides the cloud solution - including application, hardware platform, storage and other resources. Using these resources is the data controller, which in our case is an entity who has access to end-users' personal data of all kinds, in large quantities. This data may have been collected from the primary end-user either through applications installed on various personal digital devices or other means that collect user-generated content like photos, videos, documents, etc.

*A Bird's Eye View of Cloud Computing*

Cloud computing platform[3] truly emerged as a consumer oriented computing paradigm in early 2000s and soon became a popular technology. Increased bandwidth and flexible infrastructure comprising of a heterogeneous offering of softwares and hardware supported the increasing use of cloud services. It promised, and delivered, a computation environment to users with varying needs which later began to be distinguished as definitive service models (see Figure 2) – software-as-a-service (SaaS), platform-as-a-service (PaaS), and information-as-a-service (IaaS). Although there is no standard taxonomy defined, each model has been described below based on the most common features covered[4,5,6]:

   a. SaaS – All the application softwares running on a cloud infrastructure are offered to users' on-demand under the SaaS model. Also refereed to as Application Service Provider

---

[3] Mell, Peter, and Timothy Grance. "The NIST Definition of Cloud Computing, SP 800-145." NIST, Sept. 2011. Web.

[4] Takabi, Hassan, James BD Joshi, and Gail-Joon Ahn. "Security and Privacy Challenges in Cloud Computing Environments." *IEEE Security & Privacy* 8, no. 6 (2010): 24-31.

[5] Bent, Kristin. "The 20 Coolest Cloud Infrastructure, IaaS Vendors Of The 2014 Cloud 100." *CRN*. N.p., 31 Jan. 2014. Web. 03 Sept. 2014.

[6] Popek, Gerald J., and Robert P. Goldberg. "Formal requirements for virtualizable third generation architectures." *Communications of the ACM* 17, no. 7 (1974): 412-421.

(ASP), its end-user gets access to these applications (apps for short) via a thin client or web-based interface on the user device. Some of the key providers are IBM, Salesforce, Oracle, and Microsoft. Cisco is steadily making inroads into application centric infrastructure for simplified software deployment.

b. PaaS – Platform for building software applications is provided as a part of this service model. The developer, however, does not have access to underlying cloud services that may be modified using the interface. Microsoft's Azure is an example of this.

c. IaaS – In this model, computing resources such as processing, storage, and networks are provided to the user such that modifications can be made at the operating system and application level. Amazon Web Services (AWS) is the IaaS market leader with massive computational resources, aggressive pricing and an expansive product line.

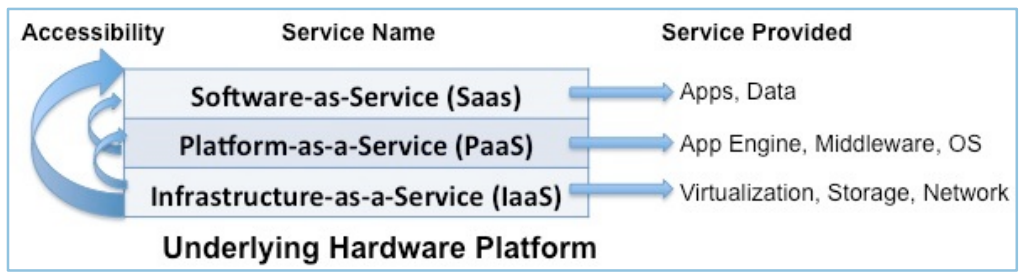

Figure 2 - Cloud service models and services provided

The vastly different needs covered by these service models are actuated by virtualization which involves creating a virtual, efficient, isolated, duplicate version of cloud resources divided into multiple execution environments. This abstraction of resources is best made possible by technologies like hypervisor that create and run on virtual machines[7]. Other softwares such as Apache Hadoop allow large scale processing of data sets with a parallel, distributed algorithm on a cluster[8].

---

[7] Seshadri, Arvind, Mark Luk, Ning Qu, and Adrian Perrig. "SecVisor: A tiny hypervisor to provide lifetime kernel code integrity for commodity OSes." *ACM SIGOPS Operating Systems Review* 41, no. 6 (2007): 335-350.
[8] Hadoop, Apache. "Hadoop." *2009-03-06]. http://hadoop. apache. org* (2009).

The "anywhere, anytime" capability of the Internet ensures a truly global solution for the cloud. Its infrastructure can be deployed using one of the three models that are described below[9]:

a. Private – The cloud provider is the only user of the infrastructure. Organization users have exclusive access to resources, which are located within the premises (physical or virtual) of the company.

b. Public – A single organization provides multiple resources to multiple consumers, which is accessed via web-services over the Internet. The overall system is located on-site or off-site which a third party provider may manage.

c. Hybrid – It is a composition of two or more internal and external cloud providers that although are independent of each other, are bound by technologies that enable inter operability of data and applications.

*Privacy: The Fundamentals*

Privacy means many things to many people. As such, this social issue is often surrounded by debates asking: *what is privacy* and *what does it mean.* These questions are informed by different philosophical approaches and theoretical views of privacy's value within a society. The most prominent of these is grounded in U.S. liberal political theory, which places the liberal self as one with the "capacity for rational deliberation and choice".[10] The liberal concept is further broken into two paradigms: positive and negative freedoms.[11] The negative paradigm is best understood as an exercise of personal choice, which is manifested in the U.S. commitment to "notice-and-choice" information practices.[12] Cohen denotes this approach as "principally defensive and ameliorative."[13] Allen, on the other hand, posits that privacy enables positive liberties such that one is free from "unwanted disclosures, publicity and loss of control of personality".[14]

---

[9] Mell, Peter ,*The NIST Definition of Cloud Computing*, 3.
[10] Julie E. Cohen, "What Privacy Is For," in *Harvard Law Review* (Vol. 126, 2013), 3.
[11] Ibid., 3.
[12] Ibid., 3.
[13] Ibid., 3.
[14] Anita L. Allen, "Coercing Privacy"*, William & Mary Law Review,* (40; 1999), 752.

From these concepts, multiple theories have emerged that explain privacy in relation to: autonomy, personhood, secrecy, liberty, etc. Warren and Brandeis, in their seminal article, *The Right to Privacy,* set forth the notion of privacy as a "right to be let alone".[15] In addition, their article outlined support for enforcement through use of tort damages, which has since heavily influenced case law. Others view privacy as accessibility to information about another person;[16] while, some refer to the harms necessary to understand privacy violations.[17] In 1960, William Prosser's article, *Privacy*,[18] analyzed hundreds of privacy cases, which informed the development of four categories within privacy tort law. These include: "1) intrusion upon the plaintiff's seclusion or solitude, or into his private affairs; 2) public disclosure of embarrassing private facts about the plaintiff; 3) publicity which places the plaintiff in a false light in the public eye; and, 4) appropriation, for the defendant's advantage, of the plaintiff's name or likeness".[19] While this work has served as a cornerstone for most modern privacy scholars, critiques of its rigidness and utility in context of information technologies have some questioning the need for revisions. More recent, Cohen offers a more post-liberal view that emphasizes the importance of recognizing a spectrum of relational and emerging subjectivity within a theory of privacy.[20]

In 1973, the U.S. Department of Health, Education, and Welfare introduced the Fair Information Practices (FIPs)[21]. This set of defined principles has since helped inform how to evaluate and design systems that may impact individual privacy rights. This framework includes guidance on transparency, individual participation, purpose specification, data minimization, use limitation, data quality and integrity, security and accountability & auditing.[22] As Schwartz notes, whenever information refers to an *identified* person, all FIPs principles should be applied - an argument he later complicates with the introduction of *identifiable* data.[23] International adoption of FIPs is apparent within certain areas of the EU's data protection plan, which include (but is not limited

---

[15] Samuel D. Warren and Louis D. Brandeis, *The Right to Privacy*, Harvard Law Review, (4; 1890).
[16] Allen, *Coercing Privacy,* 4.
[17] Daniel J. Solove. *The Digital Person*: Technology and Privacy in the Information Age, (2007).
[18] William L. Prosser, *Privacy, California Law Review,* (48; 1960), 389.
[19] Ibid.
[20] Cohen, *What is Privacy For*, 5.
[21] U.S. Department of Health, Educ. & Welfare, Records, Computers and the Rights of Citizens (50; 1973).
[22] National Institute of Standards. "National Strategy for Trusted Identities in Cyberspace: Appendix A - Fair Information Practice Principles (FIPPs)". <http://www.nist.gov/nstic/NSTIC-FIPPs.pdf> accessed 29 Aug. 2014.
[23] Paul M. Schwartz. "Information Privacy in the Cloud". Univ. of Penn. Law Review. vol. 161 at 1654.

to) the presence of an independent data protection authority and limits on automated decision making.[24]

In the U.S., the legal regime assumes various, complex approaches to address privacy concerns. It should be noted that unlike the EU, the United States does not have an omnibus information privacy statute. Instead, legal instruments, such as torts, statutes and case law, are instantiated on the state, federal, and international levels. On the constitutional level, privacy issues challenge protections afforded by the First and Fourth Amendments. In particular, critics of regulating data collection cite that such policies interfere and impede information flows, thus conflicting with the First Amendment.[25] For example, in *Sorrell v. IMS Health Inc.* (2011), the Supreme Court struck down (in a 6 to 3 decision) Vermont's prescription law[26]. The court held that the Vermont statute, which bars disclosure of prescription of data for marketing purposes, violated the free speech rights of the data-mining firms. The court determined that the prescriber-identifiable data was not fully protected speech, but instead, commercial speech; and, therefore, could not be restricted based on the Central Hudson scrutiny test.[27]

In cases of the Fourth Amendment, privacy issues arise in defining a "reasonable expectation of privacy"[28] and the need for a warrant to protect against unreasonable search and seizure by the government. Most notably, the Electronic Communications Protection Act (ECPA)[29] has recently resurfaced in Congress due in large part to increasing criticism of its outdated and insufficient alignment with modern technologies.[30] Comprised of three separate federal statues - the Stored Communications Act, the Pen Register Statute, and the Wiretap Act - its original intent was to expand Fourth Amendment protections in light of emerging computer technologies, like e-mail. Unfortunately, technology has evolved substantially over the last decade making the Act irrelevant and ill suited as a governing protocol. We discuss further implications of the ECPA as it relates to cloud services in the next section.

---

[24] Ibid., 1636.
[25] Solveig Singleton, Cato Policy Analysis No. 295. "Privacy as Censorship:A Skeptical View of Proposals to Regulate Privacy in the Private Sector." <http://www.cato.org/pub_display.php?pub_id=1154>, last accessed 28 Aug. 2014.
[26] "Act Relating to Increasing Transparency of Prescription Drug Pricing and Information", Vt. Stat. Ann. tit. 18, § 4631 (2007), ("Vermont Statute").
[27] *Central Hudson Gas & Electric Corp. v. Public Service Comm. of New York*, 447 U.S. 557 (1980). (commercial speech can be limited if: 1) truthful and non-misleading; 2) is in support of a substantial government interest; 3) directly advances the government interest asserted; and 4) is not more extensive than necessary to serve that interest).
[28] *Katz v. United States,* 389 U.S. 347 (1967).
[29] Electronic Commc'n Privacy Act of 1986, Pub. L. No. 99-508, 100 Stat. 1848 (codified at 18 U.S.C. §§ 2510—2522, 2701—2712 (2006)).
[30] Center for Democracy & Technology. "Updating ECPA." <https://cdt.org/campaign/updating-ecpa/>. last accessed 28 Aug. 2014.

Additional federal consumer privacy laws focus narrowly on types of records within specific industries.[31] The Fair Credit Reporting Act (FCRA)[32] in 1970 and the Family Educational Rights and Privacy Act (FERPA)[33] in 1974 were among the earliest federal statutes introduced in the United States. Since then, privacy laws have followed a traditional model that is reactive and measured "in terms of the value of preventing harm to an individual".[34] Examples include the Children's Online Privacy Protection Act (COPPA); the Health Information Portability and Accessibility Act (HIPAA); and the Gramm-Leach-Bliley Act (GLBA). These federal statues echo the FIPs framework in identifying potential risks for personally identifiable information (PII). Take for example COPPA (effective April 2000), which imposes limitations on the types of information that maybe collected from children under the age of thirteen. Additional provisions include parental notification and consent; the availability of a clear and detailed privacy policy by the Web site operator; and an opportunity for a parent to review any information collected by the website about his/her child.[35] Proponents, like EPIC, urged its adoption as a safeguard to prevent tracking and profiling of children online. Jeff Chester, executive director of the Center for Digital Democracy commented of the policy,

> "This proposal balances the need to protect the privacy of children, ensure parental involvement, and promotes the growth of kid-oriented online media. At a time when our children spend much of their daily lives online and are always connected to the Internet via games, cell phones and other devices, parents should thank the FTC for acting responsibly on behalf of children."[36]

---

[31] Jay P. Kesan, Carol M. Hayes, and Massoda N. Bashir, "Information Privacy and Data Control in Cloud Computing: Consumers, Privacy Preferences, and Market Efficiency". *Washington & Lee Law Review,* (70; 2013), http://http://scholarlycommons.law.wlu.edu/wlulr/vol170/iss1/6 , last accessed 27 Aug. 2014, 365.
[32] 15 U.S.C. § 168b (2006).
[33] 20 U.S.C. § 1232g; 34 CFR Part 99 (2006).
[34] Daniel J. Solove, "Identity Theft, Privacy and the Architecture of Vulnerability", *Hastings Law Journal*, (54; 2003), 5.
[35] EPIC. "COPPA's Provisions". http://epic.org/privacy/kids/, last accessed 28 Aug. 2014.

However, COPPA is not without its critics. Specifically, the provision wherein an operator must obtain veritable parental consent was seen by many as a "costly, cumbersome, and inadequate in protecting personal information".[37] Others go further criticizing COPPA's limitations on children's right to freedom of speech and self-expression. Of late, the FTC has made significant strides in enforcing COPPA by imposing large fines against companies like Disney's Playdom (fined $3 million)[38]; Xanga (fined $1 million)[39]; and Path, a mobile social networking app (settled for $800,0000).[40]

State-level statues to protect consumer data also run the gamut. In some cases, state level protections of privacy are codified within state constitutions, which expressly recognize privacy as a right (e.g. Alaska, Arizona, California, Hawaii, Illinois, Louisiana, Montana, South Carolina and Washington).[41] According to the National Conference of State Legislatures (NCSL), at least 30 states have enacted laws that specify how entities handle PII collected by businesses and governments. These laws include requirements on how to destroy and dispose of PII as well as security breach notification laws and identity theft statutes. In 2013, California enacted the Privacy Rights for California Minors in the Digital World (SB 568)[42], permitting users under the age of 18 to delete or remove content posted online (effective January 1, 2015). According to the bill, website operators must permit a minor who is a registered user to either remove or request

---

[36] Center for Digital Democracy. Jeff Chester. "Children's Privacy Advocates Praise FTC on Proposed Safeguards to Protect Children's Information Online." 15 Sept. 2011, <http://democraticmedia.org/childrens-privacy-advocates-praise-ftc-proposed-safeguards-protect-childrens-information-online> last accessed 29 Aug 2014.

[37] EPIC, "Criticisms of COPPA". http://epic.org/privacy/kids/, last accessed 28 Aug. 2014.

[38] Chloe Albanesius, "Disney's Playdom Fined $3 Million for Violating Kids' Privacy". *PC Magazine*. 16 May 2011.< http://www.pcmag.com/article2/0,2817,2385444,00.asp> last accessed 28 Aug 2014.

[39] Pete Cashmore. "Xanga Fined $1 Million for Violating Children's Privacy". *website:* mashable.com < http://mashable.com/2006/09/07/xanga-fined-1-million-for-violating-childrens-privacy/>. 7 Sept. 2006, last accessed 27 Aug. 2014.

[40] FTC. "Path Social Network App Settles FTC charge it deceived consumers and improperly collected personal information from users' mobile address books". 1 Feb 2013. < http://www.ftc.gov/news-events/press-releases/2013/02/path-social-networking-app-settles-ftc-charges-it-deceived> last accessed 28 Aug. 2014.

[41] National Conference of State Legislatures. "Privacy Protections in State Constitutions". 11 December 2013. <http://www.ncsl.org/research/telecommunications-and-information-technology/privacy-protections-in-state-constitutions.aspx>. last accessed 26 Aug. 2014.

[42] California Law. Senate Bill-568; Chapter 336

removal of information posted on their site under the "eraser" provision outlined in Section 22581.[43]

Given the amount of regulatory provisions available, one could conservatively argue that privacy issues in the U.S. are diminishing. However, despite these numerous legal layers, information privacy remains at risk of being compromised.[44] Schwartz notes that such shortcomings are a result of the law's static nature and inadequate incentives for the multiple parties who manage and store personal data to provide appropriate security and privacy protections.[45] We go further to suggest that the current use of sectoral laws and its narrowly applied approach leaves significant gaps in regulating highly distributed, modular technologies, like cloud services.[46]

*Complicating Privacy Policies and the Cloud*

As we've outlined above, state, federal and international information privacy policies vary on many different levels. Yet, they all share in the difficulty of keeping pace with technology advancements. Fast forward and legal issues of regulating cloud services are already emerging. Kesan, et. al distill current legal issues into two categories in context of cloud computing: data use and procedural issues.[47] Data use includes (but is not limited to) scraping or mining of data, use of public or private information, and the transfer of data. Kesan notes that in most data collection instances, privacy protection is unclear due to unsettled notions of whether it should prioritize the quantity of data or types of data.[48] Schwartz pushes further suggesting that changes in personal data processing have challenged traditional notions of jurisdiction, definitions of personal identifiable information, and contract law.[49] He identifies three areas of change in personal data processing due to cloud technologies: 1) nature of information processing as increasingly international; 2) multidirectional nature of modern data; 3) process-oriented

---

[43] Ibid.
[44] Identity Theft Resource. 2013. (reported nearly 58 million records were compromised)
[45] Schwartz, *Information Privacy and the Cloud,* 1624.
[46] Ibid,, 1649.
[47] Kesan, *Information Privacy and Data Control*, 365.
[48] Ibid*.,* 367
[49] Schwartz, *Information Privacy and the Cloud,* 1628-1632.

management approach, which outsources computing processes in exchange for specialization of service.[50]

As an example of the tensions between the current legal regime and cloud services, we turn our attention back to ECPA, specifically the Stored Computer Act (SCA) section. When the U.S. Congress enacted ECPA and SCA in 1986, it did so notably within the context of its own technological perspicacity. Kattan summarizes,

> "as originally enacted, the SCA attempted to balance the interests of law enforcement against individual privacy rights by dictating the mechanisms by which the government could compel a particular service provider to disclose communications stored on behalf of its customers."[51]

Sections §2702 and §2703 of the SCA specifically address when and why providers may voluntarily disclose information to others and ways in which the government may compel a provider to disclose information, respectively.[52] The SCA also identifies two types of services: electronic communication services (ECS) and remote computing services (RCS).[53] This differentiator is key in determining whether certain communications are in "electronic storage" versus just in "storage".[54] Orin Kerr offers three reasons that challenge the notion of why strong privacy protections online may not extend to the "virtual homes" in cyberspace.[55] First, he speaks directly to expectations of privacy online and the role of third-parties like Internet Service Providers (ISPs); second, pertains to the rules governing grand jury subpoenas; and third, recognizes ISPs as private actors, which means that strong protections are not extended under the Fourth Amendment.[56] Whether cloud service providers (CSPs) presumably fall under the RCS category or are identified as private actors (like ISPs) remains unclear under the current Act and its language. For e-mail service, which may be "stored" on the cloud, the government would need only a subpoena to compel the sharing of such information. Thus, organizations like Digital Due Process have urged Congress to reform ECPA and SCA in order to preserve civil liberties asking that probable cause be established and a warrant issued prior to government intrusion.

---

[50] Ibid.
[51] Ilana R. Kattan, "Cloudy Privacy Protections: Why the Stored Communications Act Fails to Protect the Privacy Communications Stored in the Cloud", *Vanderbilt Journal of Ent. & Tech.* (617; 2011).
[52] Ibid., 401.
[53] *Stored Communications Act.* 18 U.S.C. § § 2703(a)–(b).(2006)
[54] Kesan, *Information Privacy and Data Control,* 403.
[55] Orin S. Kerr, "A User's Guide to the Stored Communications Act, and a Legislator's Guide to Amending It", *George Washington Law Review*, (72; 2004), 3.
[56] Ibid., 4-5.

More perplexing legal issues surface when data is transferred across servers, state lines, and internationally. Kesan writes that the question over jurisdiction of data on the cloud surfaces due to "(1) the lack of borders in cyberspace; and (2) the vast differences between privacy laws in different locations".[57] In the United States, information privacy law does not provide government officials with the authority to block international transfers of personal information. It also does not offer any laws to regulate the processing of information unless specifically forbidden by law or regulated through other parameters.[58] Because of these jurisdictional challenges, the Terms of Service agreements between CSPs and their customers carry the burden of outlining where the data is stored and which laws apply.[59]

To illustrate the complexity of jurisdiction in context of cloud service providers, we turn to the ongoing litigation over whether Microsoft must comply with a warrant authorizing the search and seizure of email accounts hosted by the company.[60] Since 2013, Microsoft has objected to the warrant citing that the U.S. courts are not authorized to issue warrants for "extraterritorial searches". Defined in *Morrison v. National Australian Bank Ltd.*, the doctrine holds that "legislation of Congress, unless a contrary intent appears, is meant to apply only within the territorial jurisdiction of the United States."[61] In this case, Microsoft's argument hinges on the physical location of the data (stored in Ireland); while the government's rebuttal reasons that Microsoft is subject to U.S. jurisdiction and, therefore, where the data is stored is irrelevant. Recently affirmed by Judge Loretta A. Preska of the U.S. District Court for the Southern District of New York, the decision highlights unresolved tensions over jurisdiction due to conflicting international privacy law and technology outpacing policymaking. Moreover, as Kuner astutely points out, this particular discourse lacks any insight for potential safeguards for the consumers and users of services, like Microsoft's email application.[62]

---

[57] Ibid., 368-369.
[58] Schwartz, *Information Privacy and the Cloud*, 1636-1637.
[59] Kesan, *Information Privacy and Data Control,* 369.
[60] Christopher Kuner. "U.S. Warrants for Overseas Data Trample Foreign Privacy Laws". MIT Technology Review. 22 Aug. 2014 <http://www.technologyreview.com/view/530316/us-warrants-for-overseas-data-trample-foreign-privacy-law>, last accessed 26 Aug 2014.
[61] *Morrison v. Nat'l Austl. Bank Ltd.,* 561 U.S. 247, 248 (2010)

As cloud services grow, it is evident that current legal standards and regulations will need to be reformed. For some, to do so may require both legislative and FCC action.[63] Others suggest increased transparency by cloud providers as one solution. More importantly, such reform will require advancement in our conceptualization of privacy that does not exclude context and structure of contemporary data management flows. To address this, we offer additional perspectives in the upcoming section that move towards a fuller understanding of privacy issues within current cloud offering.

**PUTTING IT TOGETHER: SECURITY AND PRIVACY CONCERNS IN THE CLOUD**

In this section, we discuss security and privacy concerns that have significantly affected the deployment of cloud platform.

*Unauthorized Data Sharing*

Multi-tenancy is a part of the public or community cloud offering with the ability to roll out services to multiple users simultaneously. This supports reduced overhead and higher availability of applications for the provider through solutions management. As an instance, the ZFS storage[64] capabilities along with hypervisors, offers customized solutions down to choosing the software version, thus encouraging a modular and parallel approach. Takabi et el., further elaborate that among the unique features of the cloud, is its ability to manage resource utilization efficiently by offering a partitioned virtualized space for every customer subscribed to the service. This multi-tenancy is partially responsible for bringing the overall cost of the infrastructure down, which is why it is a great provision on behalf of the provider.

At one point during the cloud's market growth, multi-tenancy was driven by technology available at the time. Multi-tenancy architecture was accessible at the application level. Since then, several large companies have driven a major innovation wave in the cloud space leading to mature technology in the hardware, software, and virtualization space. Hence, multi-tenancy in the same storage space is no more a means of true cost reduction. An upgrade from this solution is, however, not a priority with some of the companies, including Oracle[65]. The need to move to anything has been found unwarranted and opposing business value. While saving enormous amount of money, time and human resource in creating the infrastructure from scratch is clearly an attractive offer, profit making is still the primary goal. However, prioritizing profits over the quality of product wouldn't allow sustainability for too long. As a customer, allowing ones data to reside on a multi-tenant platform poses several privacy and security related challenges. We discuss some of these challenges shortly.

---

[62] Kuner, *U.S. Warrants.*
[63] Kevin Werbach, "The Network Utility", *Duke Law Journal* 1761 (60; 2011).
[64] "Oracle ZFS Storage Software." Oracle. Accessed September 3, 2014.

Another big privacy concern is data being accessed by the service provider itself. The rationale is straightforward; there is a huge demand for data of all kinds in the Internet community. Many companies earn their livelihood by analyzing this data and selling it to interested parties at profitable rates. Platform owners usually have access and control over data inhabiting in any part of the platform, and they may take advantage of being in this unique position. If big-data analytics has given data controllers the power to extract interesting analysis from the datasets, it has also increased avenues of privacy violations. The rationale behind using cloud as a service comes understandably from reduced setup and operational costs, increased computational performance, elastic scalability, etc. This has been made possible by various service models including software (SaaS), platform (PaaS), and infrastructure (IaaS) that are offered through convenient and affordable pricing models.

The current technological barriers focus only on providing security measures in the conventional sense. Some of these are described in the following section.

*Cloud Platform Related Attacks*

CSPs[66] are the current favorites of cyber criminals, and we can expect to see more sophisticated attacks emerge in the future. They attract cybercriminals just as robbers are attracted to banks. A database located in the cloud is similar to an information bank with many customers, and cybercriminals are interested in using this data malevolently or in other unauthorized ways.

---

[65] Perry, Kathryn. "Multi-Tenancy and Other Useless Discussions." Oracle Applications Blog. July 12, 2012. Accessed September 3, 2014.

[66] "McAfee® Labs 2014 Threats Predictions." January 1, 2014. Accessed September 3, 2014.

Therefore, attacking a cloud service provider exposes entirely new and unconventional family of attack surfaces on the platform. These vulnerabilities and those implicit to other components of the infrastructure lead to many security gaps that adversaries can deploy in their favor. As an effort to thwart the efforts of the adversary or attacker, the security expert can take several measures.

The security personnel should understand attacks that are specifically geared towards the cloud platform. It is possible to shoulder the responsibility of securing the system on someone with experience in protecting enterprise network. There is definitely a basic skill that both should possess – the ability to configure, design and break security of systems.[67] Certifications from Cloud Security Alliance, Microsoft, Cisco, SANS and many other certification providers help a lot in getting hands-on experience by means of laboratory sessions and theory classes. These certifications usually provide satisfactory understanding of how various security systems work, the common attacks and protection mechanisms. This understanding can be applied safely to many systems and can also be upgraded as technology advances. Various blogs, threat reports by security firms and other organizations can keep one current with new threats on the horizon. Enough detail is provided about each threat and that progressively enhances the knowledge base of the evolving threats landscape. Below is an example of a malware threat seen on a security blog recently:

**MD5:** d685394675e2c8dd2355d6541d408896
**SHA1:** 19243fa1970f0029fdee224fdba320b478dfd976
**SHA256:** cadbdd7219e0600180c63aef23b752dea8417cd9a4c569026f72d4e6154184de
**SSDeep:** 49152:Jm75c3C2PeW0vbRdldjbTih9TnROynTFwl:JOc3C2PMv9dXTih6TI
**Size:** 1989144 bytes
**File type:** EXE
**Platform:** WIN32
**Entropy:** Packed
**PEID:** UPolyXv05_v6
**Company:** SafeInstall, LLC
**Created at:** 2014-07-31 19:30:33
**Analyzed on:** WindowsXP SP3 32-bit

Figure 3- Example of a worm description[68]

---

Using similar skills and a sound understanding of the cloud infrastructure, a security personal may feel sufficiently equipped against known threats. The attack scenarios do not, however, completely overlap in the two environments.

We can write countless papers if one were to cover existing and potential attacks implicit to cloud platforms. But, our goal is to take a systemic approach and provide mechanisms to protect the overall infrastructure instead of just focusing on safety measures through security related approaches. Our approach is novel because it explores pragmatic ways to inform the primary user of the system's capability to store data securely and in a measurable way. We recognize that the complexity of these attacks will increase as the systems continue to evolve and provide increased functionalities. Heterogeneity of constituent systems and softwares binding them will also increase overall attack surfaces. With this evaluation, we propose our solution in the next section.

## MAIN CONTRIBUTIONS

Cloud security implementers are in a prime position to develop privacy-preserving technologies to protect unlawful access of data. This is the foremost step to gain data controllers' trust for sharing data with CSPs. But, it does not offer a full proof solution guaranteeing privacy and security of all data. Much is dependent on the environment the solution is implemented in. A strong data encryption technique will protect the data from certain type of malicious intentions and approaches. There are other facets of illegal data access that necessitate new ways of protection.

Our thesis revolves around the idea of binding the distinct approaches of a technologist, a policy maker and a business owner into a combined, cohesive perspective through a secure service. Each dimension brings a critical assessment to the table but cannot guarantee a complete solution individually. In this section, we will evaluate each approach separately. Later, we propose a unified cohesive solution in the form of a big picture.

### *The Technologist's Approach*

Users of cloud services usually range from individual software application developers to small business owners who at least have a few thousand customers. Large corporations can also be seen taking advantage of the cloud to offload, conservatively speaking, part of their system. As mentioned in the Article earlier, cloud-computing infrastructure is still maturing and much work needs to be done on several frontiers, security and privacy of data being one of them. If proper data protection is not guaranteed for users, loss and exposure will ensue even in a protected premise.[69] In light of the data privacy related attacks we discussed in the previous section, we recommend performing a threat analysis of the overall system as a mandatory exercise. As we explain later, the mandate must be one assumed by the company itself and not instigated from a public policy side.

System changes configuration with each addition or removal of a hardware or software object. With the elasticity of resources that a cloud provides, a comparative threat analysis with each major upgrade will definitely offer key insights into gaping security holes and new privacy concerns resulting from the change. Before we discuss the performance of a security threat analysis,[70] we first need clarity of its meaning. A commonly accepted definition of threat, factors that create threat to a system, and ways of measuring different types of threat will be advantageous in delivering effective assessment results. For the purpose of this Article, we define a threat as "anything that is capable of acting in a manner resulting in harm to an asset and/or organization; for example, acts of God (weather, geological events, etc.); malicious actors; errors; failures."[71]

This clarification will eliminate a huge possibility of unforeseen incidents, as its impact has been realized and remedies planned in advance. It is often useful to define many separate threat models for a given system. Each model defines a narrow set of possible concerns to focus on. In the case of cloud services, possibilities of both security breach and un-handled privacy need to be modeled. This exercise can help to assess the probability, the potential ways of harm, the significance given to stored data, of attacks, and thus help limit customer concerns.

Threat Modeling has the potential to become an integral part of the process where privacy and security are relevant concerns. Some of the questions that need to be answered are:

---

[69] Kelley, Diana. "How Data-Centric Protection Increases Security in Cloud Computing and Virtualization." Security Curve. July 19, 2013. Accessed September 3, 2014.
[70] Lemoudden, M., Bouzza, N. Ben., El Quahidi, B., Bourget, D. *Journal of Theoretical & Applied Information Technology,* 8/20/2013, Vol. 53 Issue 2, p325-330, 6p, Database: Applied Science & Technology Source
[71] "Risk Taxonomy." The Open Group. January 1, 2009. Accessed September 3, 2014.

1. What types of attacks need to be modeled in a cloud environment?

2. What does threat modeling mean for a cloud platform?

3. What tools and technologies can be used to accomplish the modeling task?

4. Will threat modeling help in creating an insurance plan when a data breach takes place?

5. How to share the threat model with the customer in a readable format?

There are some existing tools that can be used to perform threat modeling for any asset that needs protection. Bruce Schneier developed attack trees[72], which is a way of thinking and describing security of systems and sub-systems. A list of possible attack vectors is created for the entire system that helps make decisions about how to improve security.

The usability of threat modeling can be maximized if the security architect thoroughly understands the architecture of a general public cloud platform. Although individual cloud-based platforms offer a broad spectrum of technologies and services, this in no way should hamper the impact this systemic analysis will have on appraisal of existing security gaps and hence proposed solutions.

Another proposed solution is to cohabitate data with same level of sensitivity and privacy needs. This first requires classification of the various user data into pre-defined sensitivity levels. These levels can be based on existing Federal or state defined policies, cost incurred in case of a data loss or can even be user designated. The ToS document can outline this agreement between the service provider and user. This proposition has the potential to increase trust on management of privacy that safeguards databases with critical data. More resources can be deployed specifically for these systems as opposed to system-wide implementation of stronger security. To further reduce ambiguity, privacy and security concerns around data can be defined bearing the following points in mind-

    a. Definition of data that needs protection

      – Who owns the data (individual/organization)?

      – What information does the data contain?

    b. Financial implications associated with data theft, unauthorized access or loss.

---

[72] "Attack Trees." Schneier on Security. December 1, 2009. Accessed September 3, 2014.

   c. Impact on owner of the data if a third party accesses the data in an unauthorized way?

   d. Federal or state imposed restrictions on access of that data

   e. Time period for which the sensitivity of the data is maximum

   *f.* To elaborate further, some of the examples are - Is the information related to a person's health records, contains sensitive information like social security number, passport number, driver's license or similar that can serve as unique identification of a person? Does the information pertain to banking account? More similar questions, if asked, can make the service provider informed of the appropriate services needed for the data to be stored and protected.

*Policymakers Approach*

In this Article, we examine privacy concerns and security protections of consumer data afforded by cloud technologies, public policies and internal data management practices. As we've advocated throughout the Article, a successful and secure cloud solution requires the integration of all three elements. Kesan et, al. recommend the introduction of a baseline of privacy protections that identify minimum requirements in order to protect sensitive information and a provision to identify risk of loss for online fraud.[73] They also call for strong enforcement and regulation of data control (e.g. data mobility, data withdrawal, secondary-use, etc.)[74] Echoing Kesan, we too believe that a baseline of protections would help provide consistent structure across the patchwork of U.S. privacy policies. In regards to personal data control, we remain on the fence for two reasons. First, usable technical mechanisms that enable such control on the user/customer level remain elusive. Second, studies indicate that personal control of data may not elicit rational behavior or good decision making by the user.[75] We also draw from Solove's challenge that "emerging privacy problems must be understood 'architecturally' as part of a larger social and legal structure. Consequently, protecting privacy must focus not merely on remedies and penalties but on shaping architectures."[76] Thus, it is in identifying how each element compliments and or detracts from the other that a comprehensive solution emerges.

---

[73] Kesan, *Information Privacy & Data Control*, 462.
[74] Ibid., 464.
[75] Acquisti, Alessandro, and Jens Grossklags. "Privacy and rationality in individual decision making." IEEE Security & Privacy 2 (2005): 24-30.

As indicated previously, one critical weakness of U.S. privacy law and regulation is the lack of incentive mechanisms for data controllers. Currently, data breaches or violations of specific policies (e.g. COPPA) result in monetary penalties. Yet, some have argued that these penalties do not go far enough to ensure consumer data protection. Thus, we ask whether other incentives may help generate a pro-privacy stance among such companies. One avenue may be in the proliferation and adoption of the "public-benefit corporation" status. State chartered, this status allows for corporations to allow a public benefit, in this case privacy, to be part of its charter purpose in addition to maximizing profits. The idea is to increase transparency and accountability of the company's efforts to protect said benefit to its customers and shareholders. On October 23, 2014, the "ad-free and never sell users' data" social networking startup, Ello, formalized its status as a b-corporation in what they noted was "the strongest legal terms possible".[77] This is a promising first step for other companies that seek to prioritize the protection of consumer data as part of their main service offerings. Yet, whether this will impact the business's bottom line or increase the user base remains unclear. Though, as consumers become more data literate, such distinctions should inform purchasing decisions. More importantly, such policies should encourage more transparent, open data practices. To ensure the viability of this recommendation, it is first worth conducting a cost analysis for both the state and for the company in consideration.

In addition to an alternative corporate structure, policymakers and companies need to re-evaluate their understanding of *personally identifiable information*. First, as we've outlined above, all data (particularly in the cloud) should be considered to be the primary asset. Second, a plethora of research over the past decade has significantly challenged the boundaries of de-anonymization, thus complicating any formal definition of PIIs as outdated. However, like Schwartz and Solove, we too consider wholly discarding PIIs as a misstep. Instead, as they propose in their PII 2.0 model, privacy should be considered on a "continuum that begins with no risk of identification at one end, and ends with identified individuals at the other".[78] In many ways, this allows for specific contexts and events to help determine the sensitivity of particular data. For example in medical use cases that may include third-party mobile applications, understanding the privacy spectrum for the data in use helps frame the legal safeguards that may need to be triggered.

---

[76] *See* Solove, *supra note* 5 at 2. (adapted to *proactively* explore the architecture of the cloud ecosystem as a means to restructure information privacy policies).

[77] Kasternakes, Jacob. "Ello becomes a public-benefit corporation with mandate not to sell ads". October 23, 2014. Last accessed on Oct. 27, 2014. < http://www.theverge.com/2014/10/23/7049141/ello-becomes-public-benefit-corporation-mandates-no-ads>

*From the Business Owners*

The Article concerns itself with the data management flow between two specific entities: the cloud service provider and users. For our purposes, we have limited the "user" or data-controller definition to include developers, start-ups, small business owners etc. We, therefore, assume a level of technical proficiency and skill-set from the user. As such, we urge these users - small business developers - to create and initiate the following steps (based or discussions above): *generate a threat model* and *share a blueprint in a layman format in addition to the ToS agreement.*

Above we offered the followings questions to ask as one begins their modeling: *What types of attacks need to be modeled in a cloud environment? What does threat modeling mean for a cloud platform? What tools and technologies be used to accomplish the modeling task?* and, *Will threat modeling help in creating an insurance plan when a data breach takes place?* After creating this model, we advocate that business owners share their models with their users. This would serve as a supplemental document to the ToS agreement, which outlines specifics of the data management flow. A second document should also be provided that discusses insurances against data breaches or mismanagement of data. We equate such an approach to a landlord with multiple tenets. Each tenet receives documentation of the property that highlights certain types of information, like the security of an area and liability for lost or damaged property. By adopting these practices, both CSPs and data controllers are clear about their data practices from data sharing, to reselling, to storage, to withdrawal, to deletion, etc. We believe that the onus lies on these entities to build and gain user trust; and that users should demand nothing less than this level of transparency.

---

[78] Schwartz, Paul & Solove Daniel. "PII 2.0: Privacy and a New Approach to Personal Information". (2012). < http://docs.law.gwu.edu/facweb/dsolove/files/BNA-PII-FINAL.pdf>

**CONCLUSION**

The technology section presented a technical recommendation that suggests threat modeling of the cloud platform and remedial actions as a complimentary solution to existing defense mechanisms. A layman version of the threat scenario of the cloud is suggested be shared with the user for transparency purposes. The next section then offered suggested pro-user management practices and internal polices to supplement the technology solution. This includes the development of a blueprint of threat modeling of the cloud infrastructure. The section on complicating privacy policies and the cloud discussed current legal privacy issues in relation to the cloud and highlighted shortcomings of current information privacy policies.

In light of these discussions, we turn our attention towards recommendations and additional points of interest that we promote will aide in securing personal data in the cloud. First and foremost, while technologies like strong encryption may be sufficient in protecting sensitive data, they are not the complete solution. *What happens when data is breached? Who is held accountable and liable? What happens when the government wants the data for an investigation?* Therefore, internal management policies and legal standards are needed.

In light of these discussions, we turn our attention towards recommendations and additional points of interest that we promote and will aide in securing personal data in the cloud. First and foremost, while technologies like strong encryption may be sufficient in protecting sensitive data, they are not the complete solution. *What happens when data is breached? Who is held accountable and liable? What happens when the government wants the data for an investigation?* Therefore, internal management policies and legal standards are needed. Thus far, the Article has primarily focused on the CSP to data-controller business relationship. Unfortunately, even at this level, the data management flow in the cloud is complex. As others and we have stated, multi-tenet, multi-directional flows on the cloud complicate legal protections. How then can we move forward?

From an internal management perspective, one point of interest is the latest decision by Apple, Inc. to explicitly ban its developers from reselling health data collected using its HealthKit API to advertising platforms, data brokers or information resellers.[79] Apple will, however, allow developers to share health data (with user consent) with third parties for "medical research purposes". Since its announcement, several questions have already emerged: will this encourage consumers to share more of their personal health data? How will this be appropriated internationally? For our purposes, we ask whether Apple's move helps fill in the gaps of HIPAA, which currently only bars private entities, like health providers and insurance companies, from communicating patient information to third parties.

Could Apple's decision set the tone for future technology data practices for all companies? While we wait to see if others follow suit, we encourage policy makers and civil society to continue efforts on to better structure current policies and statutes.

---

[79] Megha Kedia. "TOS update bars Apple HealthKit developers from selling personal data to advertisers". technienews.co.uk <http://www.techienews.co.uk/9717410/tos-apple-healthkit-developers-personal-data-advertisers/>, last accessed 31 Aug. 2014.